\documentclass[sigconf]{acmart}
\usepackage{caption}
\usepackage{subcaption}
\usepackage{graphicx}
\AtBeginDocument{%
  \providecommand\BibTeX{{%
    \normalfont B\kern-0.5em{\scshape i\kern-0.25em b}\kern-0.8em\TeX}}}

\copyrightyear{2024}
\acmYear{2024}
\setcopyright{rightsretained}
\acmConference[ASSETS '24]{The 26th International ACM SIGACCESS Conference
on Computers and Accessibility}{October 27--30, 2024}{St. John's, NL, Canada}
\acmBooktitle{The 26th International ACM SIGACCESS Conference on Computers
and Accessibility (ASSETS '24), October 27--30, 2024, St. John's, NL, Canada}
\acmDOI{10.1145/3663548.3688490}
\acmISBN{979-8-4007-0677-6/24/10}

\title[Hevelius Report]{Hevelius Report: Visualizing Web-Based Mobility Test Data For Clinical Decision and Learning Support}

\author{Hongjin Lin}
\orcid{0000-0001-6207-2147}
\affiliation{%
  \department{School of Engineering and Applied Sciences}
  \institution{Harvard University}
  \city{Allston}
  \state{Massachusetts}
  \country{United States}}
\email{hongjin_lin@g.harvard.edu}

\author{Tessa Han}
\orcid{0000-0001-7877-6210}
\affiliation{%
  \department{Bioinformatics and Integrative Genomics}
  \institution{Harvard University}
  \city{Cambridge}
  \state{Massachusetts}
  \country{United States}}
\email{than@g.harvard.edu}

\author{Krzysztof Z. Gajos}
\orcid{0000-0002-1897-9048}
\affiliation{%
  \department{School of Engineering and Applied Sciences}
  \institution{Harvard University}
  \city{Allston}
  \state{Massachusetts}
  \country{United States}}
\email{kgajos@g.harvard.edu}

\author{Anoopum S. Gupta}
\orcid{0000-0002-8741-0621}
\affiliation{%
  \department{Neurology}
  \institution{Massachusetts General Hospital}
  \city{Boston}
  \state{Massachusetts}
  \country{United States}}
\email{agupta@mgh.harvard.edu}

\renewcommand{\shortauthors}{Lin et al.}

\begin{abstract}
  \textit{Hevelius}, a web-based computer mouse test, measures arm movement and has been shown to accurately evaluate severity for patients with Parkinson's disease and ataxias. A \textit{Hevelius} session produces 32 numeric features, which may be hard to interpret, especially in time-constrained clinical settings. This work aims to support clinicians (and other stakeholders) in interpreting and connecting \textit{Hevelius} features to clinical concepts. Through an iterative design process, we developed a visualization tool (\textit{Hevelius Report}) that (1) abstracts six clinically relevant concepts from 32 features, (2) visualizes patient test results, and compares them to results from healthy controls and other patients, and (3) is an interactive app to meet the specific needs in different usage scenarios. Then, we conducted a preliminary user study through an online interview with three clinicians who were \textit{not} involved in the project. They expressed interest in using \textit{Hevelius Report}, especially for identifying subtle changes in their patients' mobility that are hard to capture with existing clinical tests. Future work will integrate the visualization tool into the current clinical workflow of a neurology team and conduct systematic evaluations of the tool’s usefulness, usability, and effectiveness. \textit{Hevelius Report} represents a promising solution for analyzing fine-motor test results and monitoring patients' conditions and progressions.
\end{abstract}



\begin{document}
%%
%% This command processes the author and affiliation and title
%% information and builds the first part of the formatted document.

\maketitle

\begin{figure*}[t!]
  \centering
  \includegraphics[width=\linewidth]{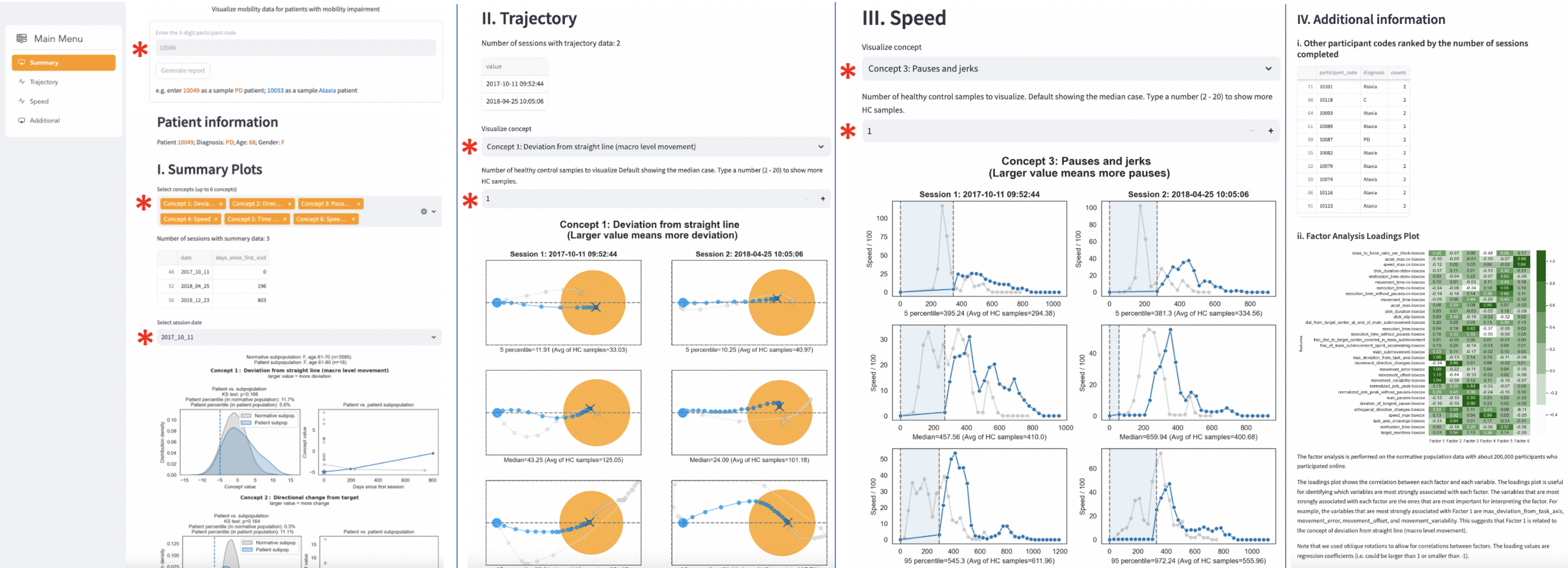}
  
  \begin{flushleft} \hspace{1.5cm} (a) Summary plots \hspace{2cm} (b) Trajectory plots \hspace{2.3cm} (c) Speed plots \hspace{2cm} (d) Additional
  \end{flushleft}
  \begin{flushleft} \hspace{15.4cm} information \end{flushleft}
  
  %\alt{Figure of the interactive visualization tool \textit{Hevelius Report}. The figure contains four sub-figures. From left to right, the first sub-figure is a screenshot of the report's first page, where a set of summary plots are shown. The second sub-figure is a screenshot of the trajectory plots. The third sub-figure is a screenshot of the speed plots. The fourth sub-figure is a screenshot of additional information, including the loading plot of factor analysis, reading instructions, and other patient unique codes ordered by the number of \textit{Hevelius} sessions they have completed (in descending order). The interface is a single continuous page with a sidebar that allows users to jump to a particular section. The red asterisks (annotations in the figure; not present in the tool) indicate opportunities for user interactions.}  
  \caption{Interactive visualization tool (\textit{Hevelius Report}) developed to help users analyze mobility test results. After the user enters a patient's unique code, the tool provides a report of the patient's test results with (a) summary plots, (b) trajectory plots, (c) speed plots, and (d) additional information on the analyses. The interface is a single continuous page with a sidebar that allows users to jump to a particular section. The red asterisks (annotations in the figure; not present in the tool) indicate opportunities for user interactions.}
  \label{fig:vistool}
\end{figure*}

\section{Introduction}

About one out of ten people in the U.S. experiences some form of mobility impairment~\cite{iezzoni2001mobility}, ranging from difficulties with gross motor skills (e.g., walking) to fine motor skills (e.g., picking up an object by hand). Clinical support, medical research, and treatment design for patients with mobility impairments rely on accurate assessments of motor ability. These assessments usually entail infrequent clinician-performed evaluations in a clinic setting. However, monitoring progression calls for more timely, frequent, and convenient assessments. \textit{Hevelius}~\cite{gajos2020computer}, a web-based computer mouse test, was developed to quantify dominant arm mobility. It has been shown to accurately measure disease severity for patients with Parkinson's disease and ataxias~\cite{gajos2020computer,Pandey2023accuracy}. A \textit{Hevelius} session spans only a few minutes and can be conducted at home~\cite{Pandey2023accuracy, Eklund2023real}. Participants are asked to perform up to 8 sets of 9 pointing movements using a computer mouse. At the end of each session, \textit{Hevelius} produces 32 features related to mouse movement profiles.

To integrate \textit{Hevelius} into clinicians' existing workflow for decision-making and research, we aim to reduce the cognitive burden required to interpret the movement data as much as possible. Taken together, our work makes the following contributions.

First, working closely with a practicing clinician-engineer, we speculated three specific obstacles clinicians may face when interpreting \textit{Hevelius} results and three potential usage scenarios. The three main obstacles are: 1) connecting the 32 fine-grained features to interpretable and clinically relevant concepts, 2) visualizing the features in digestible graphical forms, and 3) accessing and exploring the data in different user scenarios quickly and conveniently. The three main usage scenarios are: 1) reviewing a patient's data a few minutes \textit{before} a clinical appointment, 2) discussing the visualizations with patients \textit{during} the appointment (a typical clinic visit lasts 30--60 minutes), and 3) taking time to learn about the patient's data more deeply \textit{after} the appointment. 

Second, to address each of these obstacles and support the three scenarios, we developed a secure web-based visualization platform \textit{Hevelius Report} (Figure~\ref{fig:vistool}) that (1) extracts six clinically relevant concepts from 32 features with factor analysis, (2) visualizes various aspects of test results (concepts, raw movement trajectories, and speed profile) using the ``small multiple'' concept from data visualization~\cite{tufte1990envisioning}, enabling comparisons to results from healthy controls, other patients, and the patient themself over time, and (3) is interactive to meet specific needs in different user scenarios.

Third, we conducted one-hour semi-structured interviews with three additional clinicians not involved in the project to obtain preliminary user feedback on the platform. They affirmed the speculated obstacles and usage scenarios, expressing interest in learning about \textit{Hevelius} and using the \textit{Hevelius Report}. The combination of raw movement trajectories presented in the ``small multiple'' format and summary plots could effectively assist clinicians in identifying subtle changes not apparent in existing clinical tests.

Our iterative design process, visualization techniques, and findings on clinicians' needs and perspectives could be generalized to other fine-motor test results for clinical decision-support. In future work, we plan to integrate the visualization tool into the current clinical workflow of a neurology team and conduct systematic evaluations of the tool’s usefulness, usability, and effectiveness. 

\section{Related Work}

Mobility test visualizations vary based on the motor impairments and clinical environments. We focused our literature review on Parkinson's disease, which comes with some of the most common mobility problems like rigidity, slowness, and tremor~\cite{armstrong2020diagnosis}. %Prior works have developed visualizations of spiral drawing data~\cite{jusufi2014visualization}, clinical test data~\cite{hixson2023visualization}, and animated 3D avatar movement data to support clinical decision-making~\cite{piro2016analysis, synnott2010assessment, jombik2020visualization}.

\citet{jusufi2014visualization} visualized spiral drawing data of Parkinson patients collected using a telemetry touch screen device. The authors showed the raw spiral trajectory data alongside two time-series graphs showing drawing speed and deviation from the desired trajectory, suggesting a (manual) effort to reduce the raw trajectory data to two concepts: speed and position. \citet{hixson2023visualization} designed a data visualization platform to represent the Movement Disorder Society Unified Parkinson’s Disease Rating Scale Results (MDS-UPDRS). The authors displayed the MDS-UPDRS scores on human anatomical locations, which made it straightforward for clinicians to identify areas of the patient’s body that need more treatment attention. 

Prior works have also visualized the 3D movement of Parkinson's patients. For example, \citet{piro2016analysis} visualized 3D motion data using a computer vision approach to animate patients' movements while they completed a hand movement task in a clinical setting. \citet{synnott2010assessment} visualized 3D trajectory data of patients completing daily activities like raising a cup.~\citet{jombik2020visualization} displayed tremor planes in 3D to find spatial features that distinguish tremors from Parkinson's patients from healthy volunteers.  

Our work extends existing visualization efforts by providing data-driven (instead of manual) concept extraction of a web-based mobility test without requiring a specialized physical device. While our work is based on \textit{Hevelius}, our findings and visualizations are generalizable to other tests that track fine motor movement trajectories.

\section{HCI design frameworks and collaboration}
\label{sec:methods}

Our team comprises Human-Computer Interaction (HCI) scholars and a practicing neurologist-engineer, who is motivated to integrate \textit{Hevelius} into his clinical workflow. We held weekly meetings to discuss clinician needs, clinical workflows, and prototype feedback. We speculated the specific obstacles that clinicians may face when interpreting \textit{Hevelius} results:
\begin{itemize}
    \item \textbf{Obstacle 1. Clinicians may struggle to connect the 32 fine-grained features to clinical concepts.} Example features include movement offset, movement variability~\cite{keates02:cursor}, and normalized jerk~\cite{balasubramanian12:robust} (see~\cite{Pandey2023accuracy} and Appendix~\ref{app:listfeatures} for the full list). These features are too detailed and are not directly related to clinical aspects of movement disorders. This problem is exacerbated by time constraints. 
    \item \textbf{Obstacle 2. Clinicians may find interpreting \textit{Hevelius} test results in their current numerical form difficult.} Clinicians may want to visualize these numerical values in graphical forms.
    \item \textbf{Obstacle 3. Clinicians may find it difficult to access and explore the data conveniently and quickly.} The raw \textit{Hevelius} data is currently stored in CSV files. Clinicians may want to obtain the results more conveniently (i.e., without coding) and more quickly (i.e., without performing each analysis step themselves).
\end{itemize}

We also identified three main usage scenarios. First, clinicians may review patients' data \textit{a few minutes before a clinical appointment}, allowing them to grasp high-level information about their patients' arm mobility. Second, clinicians may discuss the visualization report with patients \textit{during the appointment} (a typical clinic visit lasts 30-60 minutes). Lastly, they may take more time to digest the report \textit{after the appointment}. These different usage scenarios motivate flexibility and interactivity to address distinct needs from different scenarios.

Furthermore, we co-determined the design values and revisited them throughout the process, following the \textit{Value Sensitive Design} framework~\cite{friedman1996value, ghoshal2023design}. They were 1) \textit{convey useful information} from \textit{Hevelius} for diagnosis, severity measurement, and progression tracking, 2) \textit{facilitate learning} for clinicians (and patients) about disease progression and treatment, and 3) \textit{be mindful} about how certain visualizations might make patients feel discouraged and confused. While we focus on clinicians' needs in this work, an ongoing study from our team seeks patients' perspectives to update this set of values (and our work) to better serve the patients as key stakeholders~\cite{so2024its}. %Defining the design values early on allowed us to keep our end users in mind and avoid diverging from their needs. 

Lastly, we presented multiple designs to solicit informal feedback in our weekly meetings with our clinician-engineer collaborator, following a \textit{Parallel Design} principle~\cite{tohidi2006getting, dow2011prototyping}. 
\section{Data}

We used two data sources in this work. One dataset was collected from 247,667 healthy control participants in a web-based study hosted on \textit{LabintheWild}~\cite{reinecke15:labinthewild} in 2013. We used this dataset for factor analysis to extract clinically relevant concepts. Another dataset was collected from 286 participants from a clinical study (2016--2020) conducted by the clinician in our team, with 124 of the participants diagnosed with ataxias, 61 diagnosed with Parkinson's, 59 of them serving as healthy controls, and the rest with other diagnoses. About 70 ($24\%$) participants completed more than one session. We used this dataset to develop visualizations for individual patients. Both datasets were anonymized before the authors could access them.  
\section{Factor Analysis and Data Visualizations}
\label{sec:viztool}

We describe three aspects of the visualization tool, each addressing an obstacle described in Section~\ref{sec:methods}. We further describe other solutions explored and explain why the chosen solution is optimal in Appendix~\ref{app:othersolutions}.

\textbf{First, the visualization tool extracts high-level, clinically relevant concepts.} Addressing the obstacle of interpreting fine-grained features (Obstacle 1), we performed factor analysis \cite{tavakol2020factor} on 32 \textit{Hevelius} features and obtained six higher-level concepts, using the \textit{factor\_analyzer} Python package~\cite{factor_analyzer}. We implemented the \textit{promax} rotation (an oblique rotation method) to allow the factors to be correlated since many clinical concepts, such as speed and accuracy of movement trajectories, are correlated~\cite{Watkins2018exploratory}. 

Figure~\ref{fig:factor-analysis} (Appendix~\ref{agg:vis}) represents the loading between each factor and \textit{Hevelius} feature. Features with the highest loading values for a factor are the most salient for interpreting the factor. For example, the variables that are most strongly associated with Factor 1 are the maximum deviation from the task axis, movement errors, movement offset, and movement variability. This list of variables led us to infer that Factor 1 is related to deviation from a straight line. We mapped each factor to a concept relevant to aspects of movement that clinicians are likely to assess when examining patients' motor skills. The resulting six concepts are 1) deviation from the straight line, 2) directional change from the target, 3) pauses and jerks, 4) speed, 5) time inconsistency, and 6) speed inconsistency. Two categories emerged from the six concepts: the first two are related to the computer mouse trajectory, while the latter four are related to time and speed. We are conducting additional analysis to investigate the statistical relationship between these six concepts and clinical scores like the Brief Ataxia Rating Scale (BARS)~\cite{schmahmann2009development} and Patient-Reported Outcome Measure of Ataxia (PROM-Ataxia)~\cite{Schmahmann2021prom}.

\textbf{Second, we visualized various aspects of \textit{Hevelius} test results (concepts, trajectory profile, and speed profile),} addressing the obstacle of interpreting the data in its current numerical form (Obstacle 2). First, we created visualizations to summarize information about each of the six concepts (\textbf{``summary plots''}). An example summary plot for the ``deviation from the straight line'' concept is shown in Figure~\ref{fig:summary-plot} (Appendix~\ref{agg:vis}). We overlaid the patient’s value over the distributions of comparable healthy controls (i.e., same gender and age group) and comparable patients (i.e., same diagnoses, gender, and age group) (Figure~\ref{fig:summary-plot}, left). We also plotted the patient's values and those of comparable patients over time (Figure~\ref{fig:summary-plot}, right). These visualizations enable clinicians to compare the patient to healthy controls and other patients and track the patient’s progression. 

Next, we visualized how the patient moved the computer mouse to the target (\textbf{``trajectory plots''}), using the ``small multiple'' concept from data visualization~\cite{tufte1990envisioning}. An example trajectory plot is shown in Figure~\ref{fig:trajec-plot} (Appendix~\ref{agg:vis}). For a given concept, we identified the $5th$, $50th$, and $95th$ percentile of trials of the patient (in blue) and the healthy control population (in grey). We plotted these trajectories in three rows (one row for each of the three percentiles). Each column represents one \textit{Hevelius} session. This visualization enables one to compare the mouse trajectories of the patient and those of healthy controls, see the variation in trajectories across trials at a given time point (best, median, and worst), and see how trajectories change over time. 

Furthermore, we visualized the mouse movement speed profile during each trajectory (\textbf{``speed plots''}). An example speed plot is shown in Figure~\ref{fig:speed-plot} (Appendix~\ref{agg:vis}). The speed plot shows three phases of the task: initiation (between the start of the trial and the start of mouse movement), execution (between when the patient first moves the mouse and before the patient clicks the mouse), and click duration (the start of the click to the end of the click, i.e., from mouse down to mouse up). This visualization enables one to compare the computer mouse speed of the patient and healthy controls, see the variation in speed across trials at a given time point ($5th$, $50th$, and $95th$ percentile trials), and see how movement speed changes over time. 

Through representing contrasts~\cite{marton2014necessary}, the visualizations can provide insights into a patient's mobility condition and progression compared to other populations. For example, patients with Parkinson's might move the mouse more slowly but follow a straight line. In contrast, healthy control participants might move quickly but deviate from the straight line more drastically. Healthy controls might also start the task much sooner than patients in the initiation phase, which connects with the clinical concept of movement initiation.

\textbf{Third, the visualization tool is interactive.} Upon developing a design prototype (Appendix~\ref{app:dev-ui}, Figure~\ref{fig-app:dev-ui}), we built an interactive app---\textit{Hevelius Report}---using \textit{Streamlit.io}~\cite{streamlit} to address the clinicians' need to conveniently explore \textit{Hevelius} results (Obstacle 3). When using the visualization tool, users are first asked to enter the unique code of a patient. Then, users are presented with basic patient information (diagnosis, age, gender) (Figure~\ref{fig:vistool}a), summary plots (with the option to view different dates) (Figure~\ref{fig:vistool}a), trajectory plots (Figure~\ref{fig:vistool}b), speed plots (Figure~\ref{fig:vistool}c), and additional information about the analyses (specific factor analysis results) (Figure~\ref{fig:vistool}d). Users can select a subset of concepts and set the number of health control comparisons (in grey) as they scroll through the report. 

\section{Preliminary user study}

\textbf{Recruitment and Analysis.} To elicit early feedback on the usefulness and usability of our visualization platform, we conducted a one-hour semi-structured group interview with three clinicians who treat patients with movement disorders, including Parkinson's disease. None of them were involved in the project or had prior knowledge of the system. We recruited the clinicians through personal connections. Since clinicians are highly specialized experts with many time constraints, we minimized the number of scheduling emails and agreed to meet at a time that worked for all three of them. We did not record the meeting nor collect any personal information. We only reported high-level findings without using any quotes in this paper. 

\textbf{Results.}  Overall, the clinicians affirmed the speculated obstacles and usage scenarios, expressing interest in using this visualization tool. The combination of raw movement trajectories and summary plots allowed them to identify subtle changes not available in existing clinical tests. The clinicians confirmed that the high-level concepts extracted using factor analysis were clinically relevant. The clinicians indicated that those concepts are aspects of movement they examine when assessing a patient's motor skills. They also expressed interest in further analysis of the statistical associations between the factors and established clinical measures. 

When it comes to the visualizations, clinicians held diverging preferences, highlighting the importance of interactivity. For example, regarding the ordering of data visualizations, two clinicians preferred seeing summary plots before diving into the details; one clinician preferred seeing trajectory plots first to do most of the synthesis himself before seeing the summary plots. This indicates different preferences for cognitive engagement \cite{gajos2022people}. Designing to meet human experts' diverse needs in clinical tasks is crucial for reducing over-reliance on automation and the resulting high-stake erroneous decisions \cite{goddard2012automation}.

\section{Limitations and Future Work}

One limitation of our current work is the relatively small number of end users involved. Future work will seek perspectives from more clinicians, as well as other key stakeholders like patients (e.g., to understand how they might want to see and use their own data~\cite{so2024its}). Future work also includes integrating \textit{Hevelius Report} into an examination program \textit{Neurobooth}~\cite{neurobooth} that uses comprehensive digital tests, including \textit{Hevelius}, to examine a patient's motor skills, eye movements, speech, and cognition. In the meantime, we will seek additional user perspectives on user scenarios and obstacles. Upon further improvement, we will formally evaluate the tool and its impact on clinical decision-making by conducting systemic evaluations with clinicians and other end users--with usability testing like the system usability scale (SUS) questionnaire and controlled studies. 

Another limitation is the small number of longitudinal data to visualize patients' progression of mobility over time. We believe that ongoing data collection efforts will enable better usage of the progression plot. We also emphasize the importance of recruiting a demographically and socio-economically diverse group of patients and healthy control participants to mitigate downstream bias~\cite{kamikubo2022data}. A diverse and accurate data presentation of patients will improve the population estimates of \textit{Hevelius} outcomes and facilitate more equitable clinical research and treatment. 

\begin{acks}
We thank our clinician participants for their valuable feedback. The project benefited greatly from conversations with Jianna So, Faye Xiao Hui Yang, Jakob Troidl, Zana Buçinca, Catherine Yeh, and Elena L. Glassman. We also thank Siddharth Patel for his support in obtaining the necessary data for the project. 

This work was supported in part by the National Institutes of Health under Grant No. R01 NS117826 and by the National Science Foundation under Grant No. IIS-2107391. Any opinions, findings, conclusions, or recommendations expressed in this material are those of the authors and do not necessarily reflect the views of the National Institutes of Health or the National Science Foundation.
\end{acks}

\bibliographystyle{ACM-Reference-Format}
\bibliography{references}

\clearpage
\appendix

\section{Figures of visualization}
\label{agg:vis}
\begin{figure}[h]
\centering

  \includegraphics[width=\linewidth]{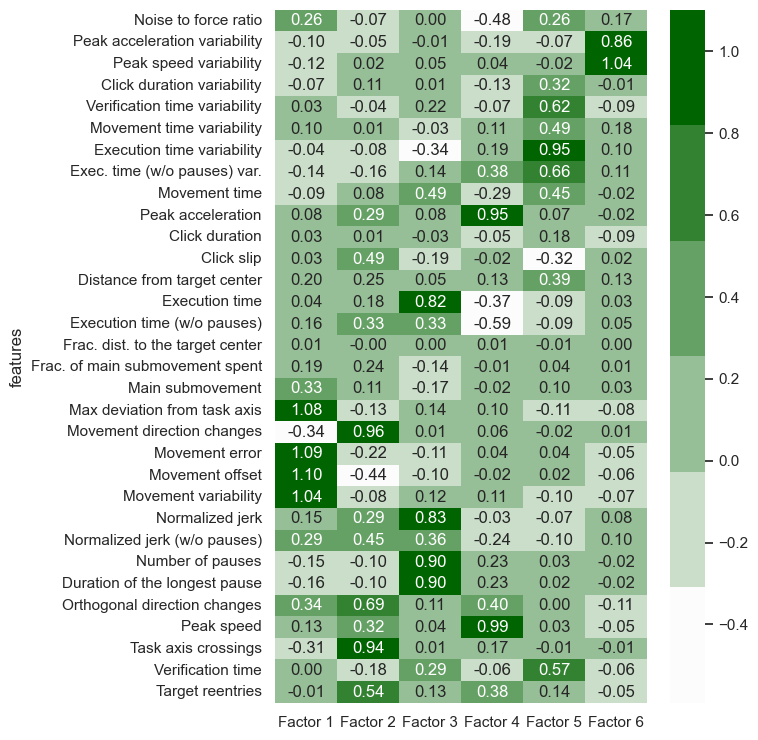}
  %\alt{A loading plot in a heatmap format. Y axis is the list of 32 \textit{Hevelius} features and x axis is the list of 6 factors. Each cell contains a numeric value that is the loading coefficient of the feature onto the factor. The higher the value, the darker the color (green) is.}
  \caption{Factor analysis results. We performed factor analysis on the \textit{Hevelius} test results from 247,667 healthy control participants and reduced the 32 \textit{Hevelius} features to 6 factors. The coefficients representing the relationship between each factor and the features are displayed above. We mapped the 6 factors to 6 concepts: 1) deviation from straight line, 2) directional change from target, 3) pauses and jerks, 4) speed, 5) time inconsistency, and 6) speed inconsistency (Factors 1-6, respectively). Discussions with clinicians indicated that these concepts are clinically relevant.}
  \label{fig:factor-analysis}
\end{figure}

\begin{figure}[h]
  \centering
  \includegraphics[width=\linewidth]{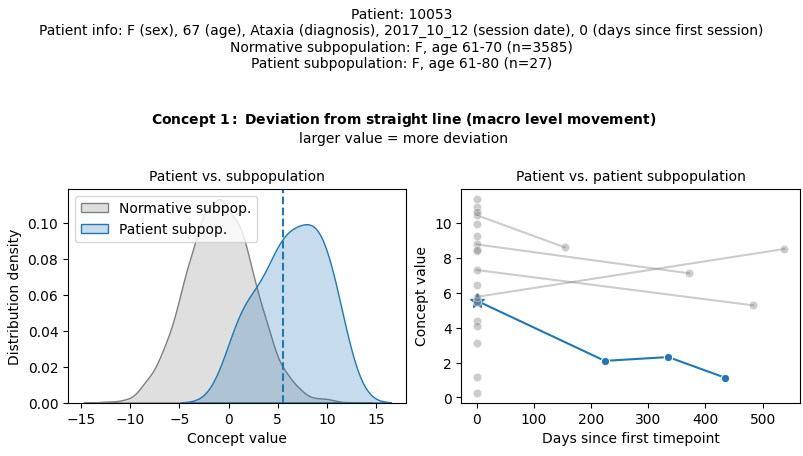}
  %\alt{Example summary plot for a Parkinson's patient for a given timepoint and concept. There are two sub-plots in this image. On the left plot, the y-axis is distribution density, and the x-axis is the value of the chosen concept (a linear combination of the 32 features based on factor analysis). The value of the patient is plotted as a dotted verticle line (blue). The distributions of comparable healthy controls (in grey) and other patients (in blue) are plotted in the background. On the right plot, the y-axis is the concept value, and the x-axis is the days since the patient's first Hevelius session. The patient's values are plotted as a blue line, and the comparable healthy control averages are plotted as grey.}
  \caption{Example summary plot for a Parkinson's patient for a given timepoint and concept. The plot on the left compares the patient's result with that of the relevant patient and healthy control sub-populations. The plot on the right shows the progression of the patient's condition over time and compares the patient's progression with that of other patients.}
  \label{fig:summary-plot}
\end{figure}

\begin{figure}[h]
\centering
  \includegraphics[width=0.9\linewidth]{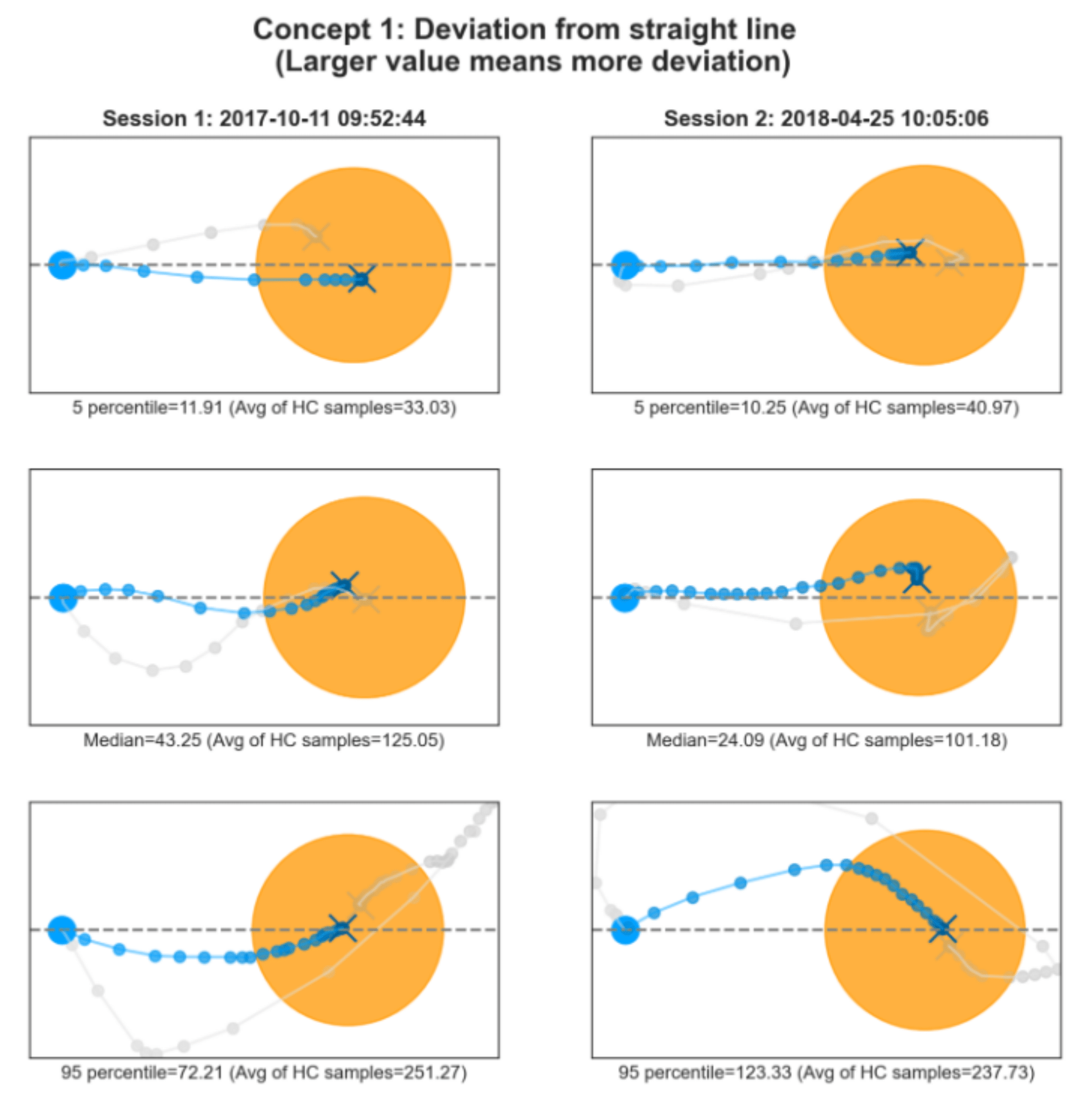}
  %\alt{Example trajectory plots for a Parkinson's patient. There are six sub-plots in this figure. Each column represents data from one Hevelius session. Each row represents data from a percentile (5th, 50th, or 95th) of the concept value. Within each sub-plot, a big orange circle on the right represents the target region of a Hevelius trial. The smaller blue dot on the left represents the starting point of the computer mouse. The dotted blue line represents the mouse's trajectory from the starting point to the target region. The grey dotted line represents a healthy control average trajectory.}
  \captionof{figure}{Example trajectory plots for a Parkinson's patient. The plot compares the mouse trajectory of the patient (blue) to example trajectories from healthy controls (gray) over two \textit{Hevelius} sessions (two columns). Users can choose to see one average healthy control trajectory or a distribution of up to 20 trajectories. The $5th$, $50th$, and $95th$ percentile trajectories are selected based on a concept (in this example, the "deviation from straight line" concept). }
  \label{fig:trajec-plot}
\end{figure}

\begin{figure}[h]
  \centering
  \includegraphics[width=0.9\linewidth]{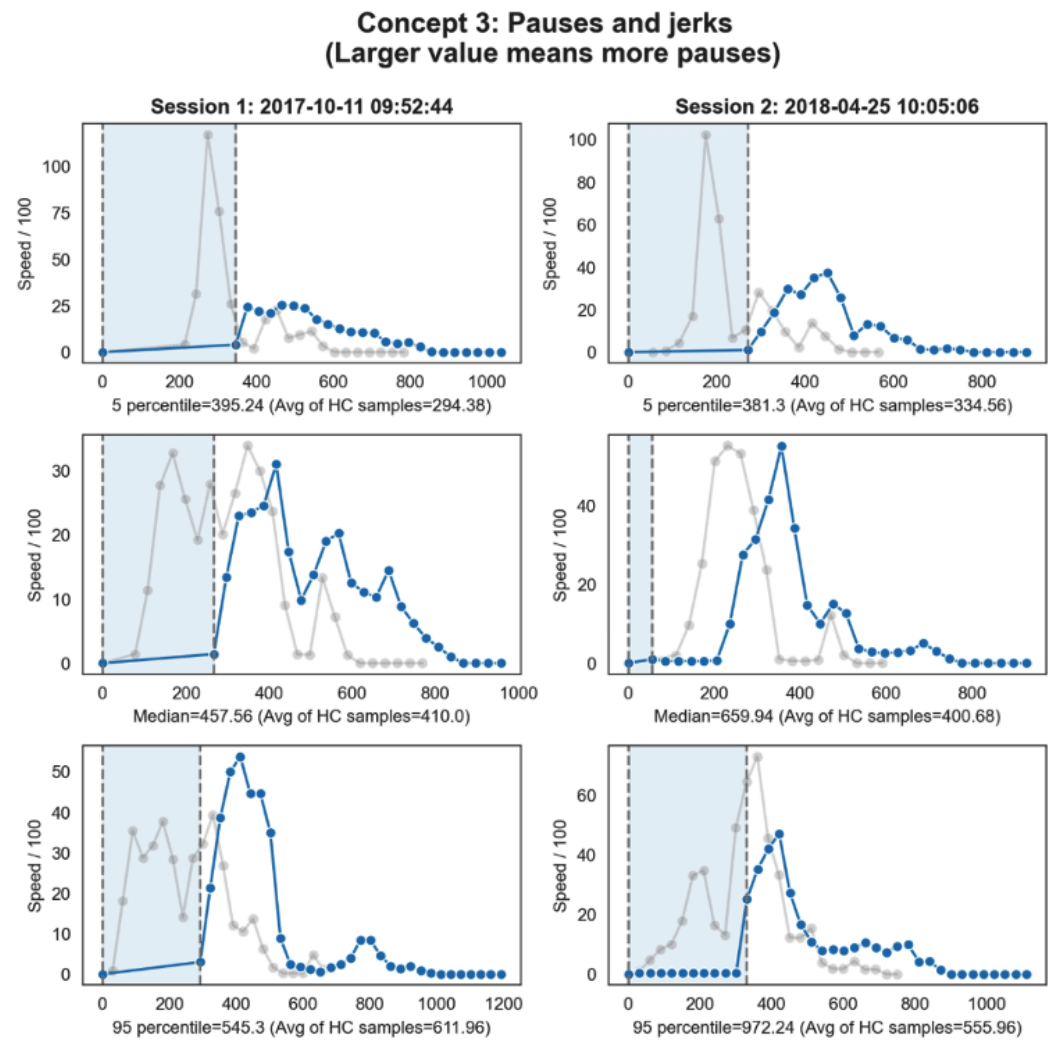}
  %\alt{Example time-speed plots for a Parkinson's patient over two \textit{Hevelius} sessions. The plot contains six sub-plots. Each column represents data from one Hevelius session. Each row represents data from a percentile (5th, 50th, or 95th) of the concept value. Within each sub-plot, the y-axis measures speed, and the x-axis is the time point of a Hevelius trial. The speed profile of the mouse movement is presented as a line. The patient's line is blue, and the healthy control average is grey. The verticle dotted lines represent when the trial begins and when the patient starts to move the mouse.}
  \captionof{figure}{Example time-speed plots for a Parkinson's patient over two \textit{Hevelius} sessions. The plot shows the speed of the mouse during the corresponding trajectories in the trajectory plot (Figure~\ref{fig:trajec-plot}). Users can choose to see one average healthy control speed line (gray) or a distribution of up to 20 speed lines. The initiation phase (time between the start of the task and the start of mouse movement) is highlighted in the light blue shaded area.}
  \label{fig:speed-plot}
\end{figure}

\clearpage
\section{Other solutions explored}
\label{app:othersolutions}
\subsection{Concept Learning}
\begin{figure}[h]
  \centering
  \includegraphics[width=\linewidth]{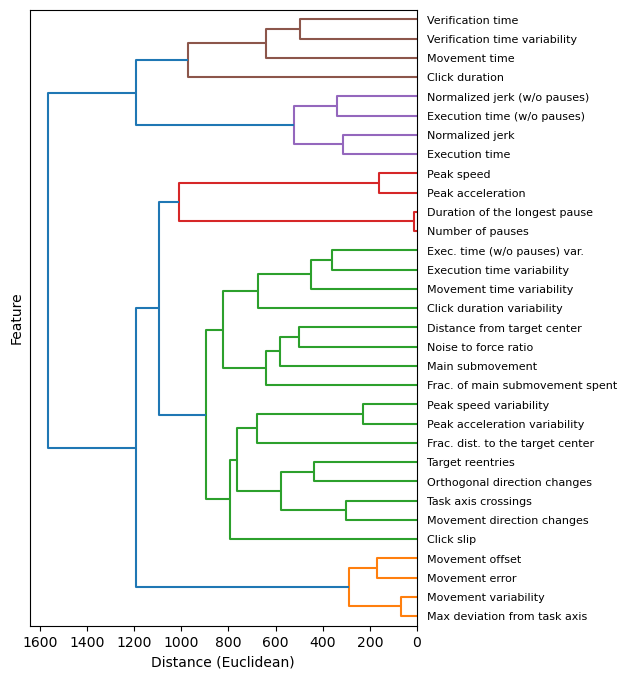}
  %\alt{A dendrogram from hierarchical clustering. The x-axis is the Euclidean distance, and the y-axis is the list of 32 Hevelius features. The features are grouped based on their Euclidean distance.}
  \caption{Hierarchical clustering. In addition to factor analysis, we also explored hierarchical clustering as another method to reduce the dimensionality of the \textit{Hevelius} features. The dendrogram from the hierarchical clustering analysis is displayed above. Although hierarchical clustering shows how features are related to each other in terms of distance in space, it is unclear how to summarize a given cluster of features mathematically. Thus, we did not end up using hierarchical clustering to reduce the dimensionality of the features.}
  \label{fig-app:hierarch-clus}
\end{figure}

In addition to factor analysis, we also explored other methods to perform dimensionality reduction, including k-means clustering \cite{hartigan1979algorithm}, t-SNE \cite{van2008visualizing}, and hierarchical clustering \cite{murtagh2012algorithms}. However, these methods had the following shortcomings. Although k-means clustering and t-SNE can reduce a large number of features into a smaller number of clusters, the relationship among features in each cluster and, thus, the meaning of each cluster is difficult to interpret. For hierarchical clustering, although the resulting dendrogram depicts how features are related to each other in terms of distance in space, it is unclear how to summarize a given cluster of features mathematically. For details, please see Figure~\ref{fig-app:hierarch-clus}. Because factor analysis not only clearly indicates the relationship between each factor and the features but also provides a way to mathematically summarize the features (i.e., each factor is a linear combination of the features), we chose to use factor analysis for the study. When performing factor analysis, we explored the possibility of reducing the 32 features to 4, 5, 6, and 7 factors. We ultimately selected 6 factors because they provided a good balance between a small number of factors explaining a large proportion of variance in the data. 

We also conducted factor analysis on patient subgroups instead of healthy control populations to understand how the concept groups differ. We found a high level of overlap among these results and decided to use the concept groups based on the healthy control populations due to their generalizability across different patient groups.

\subsection{Visualization}
\begin{figure}
\centering
\begin{subfigure}{0.7\linewidth}
  \centering
  \includegraphics[width=\linewidth]{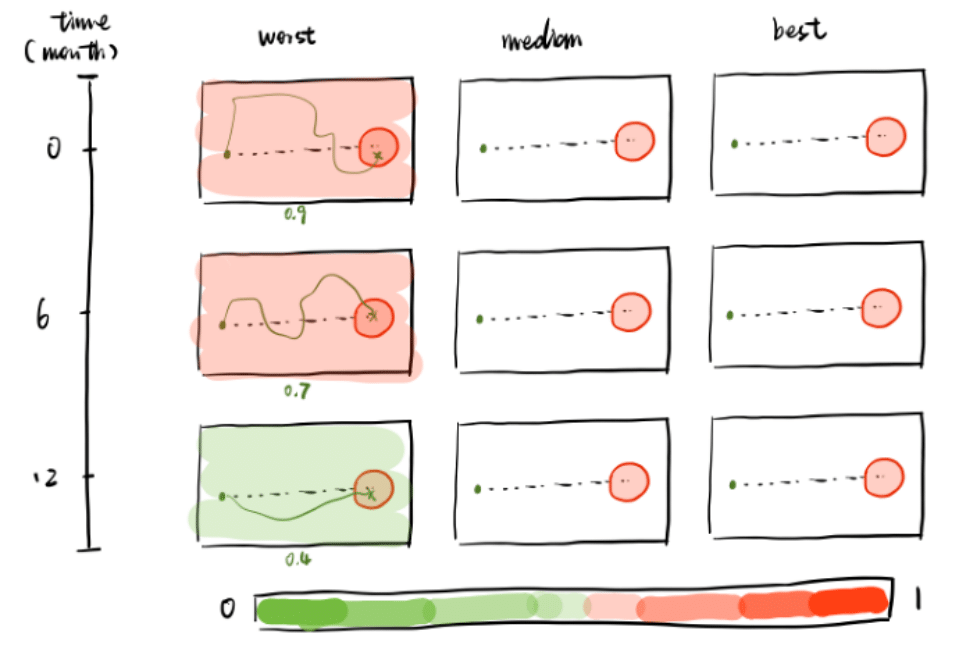}
  \caption{Design sketch}
\end{subfigure}
\begin{subfigure}{0.7\linewidth}
  \centering
  \includegraphics[width=\linewidth]{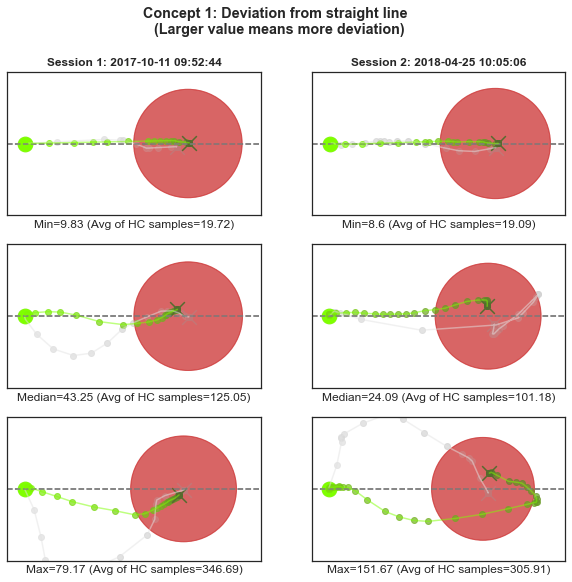}
  \caption{Earlier version}
\end{subfigure}
\begin{subfigure}{0.7\linewidth}
  \centering
  \includegraphics[width=\linewidth]{figures/fig4-trajec-plot.png}
  \caption{Latest version}
\end{subfigure}
%\alt{Comparison between two earlier versions and a final version of the trajectory plots. The caption details the changes between the three versions.}
\caption{Development of trajectory plots, from (a) a design sketch to (b) an earlier version to (c) the latest version (same as Figure~\ref{fig-app:dev-trajec-plots} in main paper). After sketching potential designs, we chose the design shown in (a) and implemented it in (b). In contrast to the earlier version (b), which visualizes the best, median, and worst trajectories, shows the average healthy control trajectory in gray, and uses a red-green color scheme, the latest version (c) visualizes the $5th$, $50th$, and $95th$ percentile trajectories, can show a single average or up to 20 healthy control trajectories in gray (user decides), and uses an orange-blue color scheme consistent with the color scheme in the summary and speed plots.}
\label{fig-app:dev-trajec-plots}
\end{figure}

\begin{figure*}
\centering
\begin{subfigure}{0.5\linewidth}
  \centering
  \includegraphics[width=0.9\linewidth]{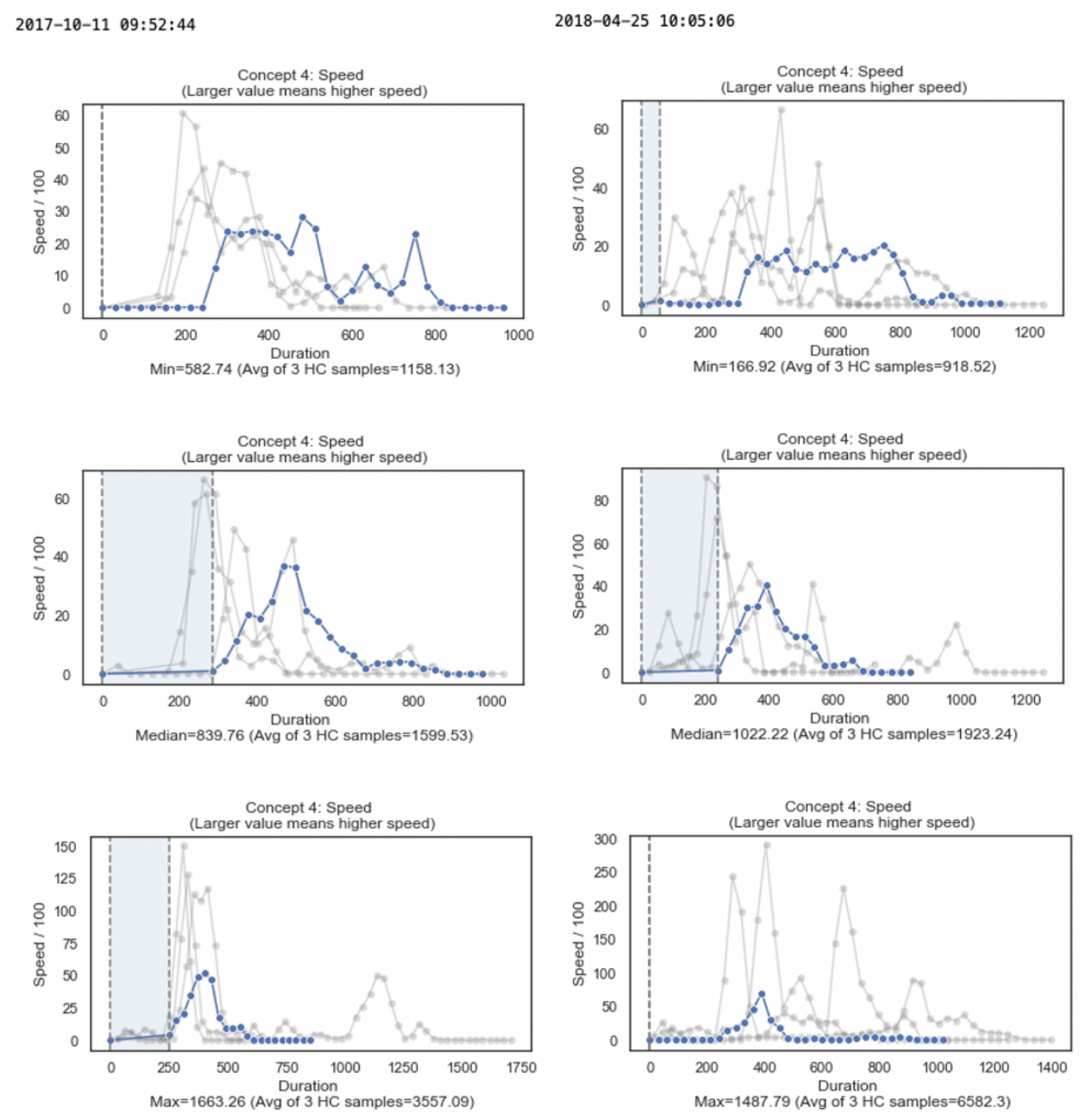}
  \caption{Earlier version}
\end{subfigure}%
\begin{subfigure}{0.5\linewidth}
  \centering
  \includegraphics[width=0.9\linewidth]{figures/fig5-speed-plot.png}
  \caption{Latest version}
\end{subfigure}
%\alt{Comparison between an earlier version and a final version of the speed plots. The caption details the changes between the two versions.}
\caption{Development of speed plots, from (a) an earlier version to (b) the latest version (same as Figure~\ref{fig-app:dev-speed-plots}. In the earlier version, users could only visualize the average speed line for a healthy control in gray. In the latest version, users can visualize a single average or up to 20 healthy control speed lines in gray.}
\label{fig-app:dev-speed-plots}
\end{figure*}

During the design of \textit{Hevelius} visualizations, one key question that came up repeatedly was how to show the distributions of the healthy control subpopulation and the patient subpopulation. For the summary plots, we obtained the distribution of values, performed kernel density estimation to obtain a smooth curve, and plotted this curve to represent the distribution. For the trajectory and speed plots, based on discussions with our clinician collaborator, we initially plotted three trajectories from the specified quantiles from healthy control participants to provide some information about the distribution of trajectories. However, discussions with other clinicians revealed that some prefer seeing a single average trajectory from the healthy control population instead. To account for these different preferences, we designed the app to allow users to specify the number of trajectories they want to see. For examples of different prototypes during the development of the visualizations, please see Appendix~\ref{app:dev-data-vis} (Figures~\ref{fig-app:dev-summary-plots},~\ref{fig-app:dev-trajec-plots}, and~\ref{fig-app:dev-speed-plots}).

Another point of discussion was whether to normalize data points to facilitate the comparison of patients' progression in the summary plots (Figure~\ref{fig:summary-plot}, right). In the plot, normalizing each patient's data with respect to results from each patient's first \textit{Hevelius} session would allow all the lines to start at the same starting point and would allow for a clearer comparison of how each patient has progressed since their first \textit{Hevelius} session. However, the unnormalized data shows large variability in the patients' starting points. This information would be lost upon normalization and may misguide the interpretation of the data, misleading one to believe that patients are more similar than they actually are. Given these concerns with normalized data, we opted to visualize the unnormalized data.

\subsection{Visualization Platform}

In addition to \textit{Streamlit.io}, we also considered building the user interface using \textit{R Shiny} and via our own website. Building our own website would allow for great flexibility, but would also entail unnecessary additional work of building various website structures from scratch. In contrast, \textit{Streamlit.io} and \textit{R Shiny} are packages that offer support in building interactive web apps, so we opted for the latter two. We chose \textit{Streamlit.io} over \textit{R Shiny} to build the user interface because \textit{Streamlit.io} integrates seamlessly with Python (the coding language used to create the data visualizations).

During the design process, we discussed the interface options, such as having different pages for each data visualization or a continuous scroll. Based on discussions with our collaborators, we opted for a continuous scroll so that users could see multiple visualizations simultaneously (if they choose to) and added a sidebar menu to jump to different sections quickly.

\newpage
\section{List of features in \textit{Hevelius} test results}
\label{app:listfeatures}
The \textit{Hevelius} test results include 3 features related to patient and session identification and 32 features related to mouse movement. The 3 features related to patient and session identification are 1) patient ID, 2) diagnosis, and 3) patient-session ID. The 32 features related to mouse movement are:

\begin{enumerate}
  \item Standard deviation (computed over all trials in a block) of the distance from the target center at the end of the first submovement divided by the mean of peak accelerations
  
  \item Peak acceleration: maximum smoothed acceleration recorded during a movement
  
  \item Peak speed: the maximum smoothed speed recorded during a movement 

  \item Click duration variability: standard deviation of click durations in a block of trials

  \item Verification time variability: standard deviation of verification times in a block of trials

  \item Movement time variability: coefficient of variation of movement times in a block of trials

  \item Execution time variability: coefficient of variation of execution times in a block of trials

  \item Execution time variability without pauses: coefficient of variation of execution times without pauses in a block of trials

  \item Movement time: complete movement time from target onset to the end of the successful click on the target

  \item Peak acceleration: maximum smoothed acceleration recorded during a movement

  \item Click duration: time between mouse button press and release during the correct click on the target

  \item Click slip: distance between the point where the mouse button was pressed down and where it was released during click on the target

  \item Distance from target center at the end of main submovement: 2D distance from the mouse pointer location at the end of the main submovement to the target center

  \item Execution time: time from the first to the last mouse movement, excluding any movement that occurred while the mouse button was pressed

  \item Execution time without pauses: like execution time, but excludes pauses of 100ms or longer

  \item Fraction of remaining distance to the target center covered in main submovement: fraction of the remaining distance along the task axis covered during the main submovement

  \item Fraction of the main submovement spent accelerating: fraction of the time from the start of the submovement to the time when acceleration reached its peak value divided by the overall duration of the submovement

  \item Main submovement: submovement with the highest peak speed

  \item Maximum deviation from task axis: maximum distance of the mouse pointer from the task axis during a movement

  \item Movement direction changes: number of times the movement component orthogonal (vertical) to the task axis changes sign

  \item Movement error: average absolute distance of the mouse pointer from the task axis

  \item Movement offset: average (non-absolute) distance of the mouse pointer from the task axis 

  \item Movement variability: standard deviation of the distance of the actual path from the task axis

  \item Normalized jerk: dimensionless measure computed as normalized jerk based on~\citet{Pandey2023accuracy}

  \item Normalized jerk without pauses: like normalized jerk, but excludes parts of the movement when the mouse pointer was paused for 100ms or longer

  \item Number of pauses: number of pauses of 100ms or longer

  \item Duration of the longest pause: duration of the longest pause of 100ms or longer

  \item Orthogonal direction changes: number of times the movement component parallel to the task axis changes sign

  \item Peak speed: maximum smoothed speed recorded during a movement

  \item Task axis crossings: number of times the mouse pointer crossed the task axis during the movement

  \item Verification time: time interval between the end of a movement inside a target and the beginning of the click (i.e., the time when the mouse button was pressed)

  \item Target re-entries: number of times the mouse pointer leaves the target and then re-enters it before the start of the click
\end{enumerate}

\clearpage
\section{Development of data visualizations}
\label{app:dev-data-vis}

\begin{figure}[h]
\centering
\begin{subfigure}{\linewidth}
  \centering
  \includegraphics[width=\linewidth]{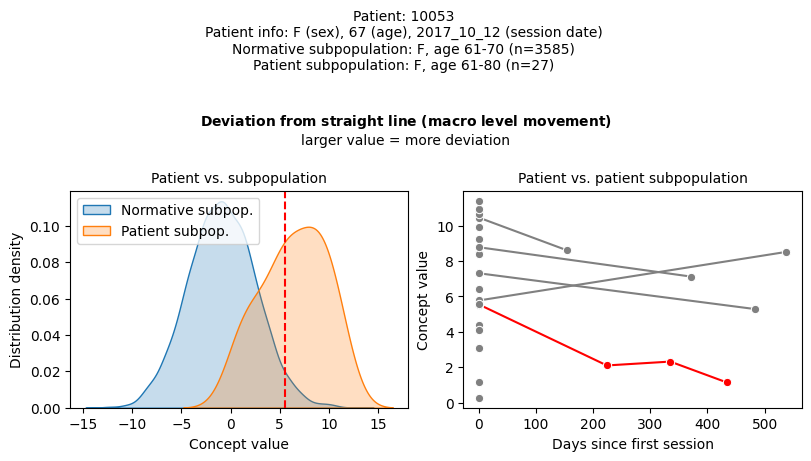}
  \caption{Earlier version}
\end{subfigure}
\begin{subfigure}{\linewidth}
  \centering
  \includegraphics[width=\linewidth]{figures/fig3-summary-plot.png}
  \caption{Latest version}
\end{subfigure}
%\alt{Comparison between an earlier version and a final version of the summary plots. The caption details the changes between the two versions.}
\caption{Development of summary plots, from (a) an earlier version to (b) the latest version (same as Figure~\ref{fig:summary-plot} in main paper). In the latest version, we adjusted the color scheme for consistency with trajectory and speed plots, added a star indicating the visualized timepoint in the progression plot, statistically tested the difference between the patient population and control population, and noted the patient's percentile with respect to the patient population and healthy control population.}
\label{fig-app:dev-summary-plots}
\end{figure}

\clearpage
\section{Development of user interface}
\label{app:dev-ui}

\begin{figure}[h]
  \centering
  \includegraphics[width=\linewidth]{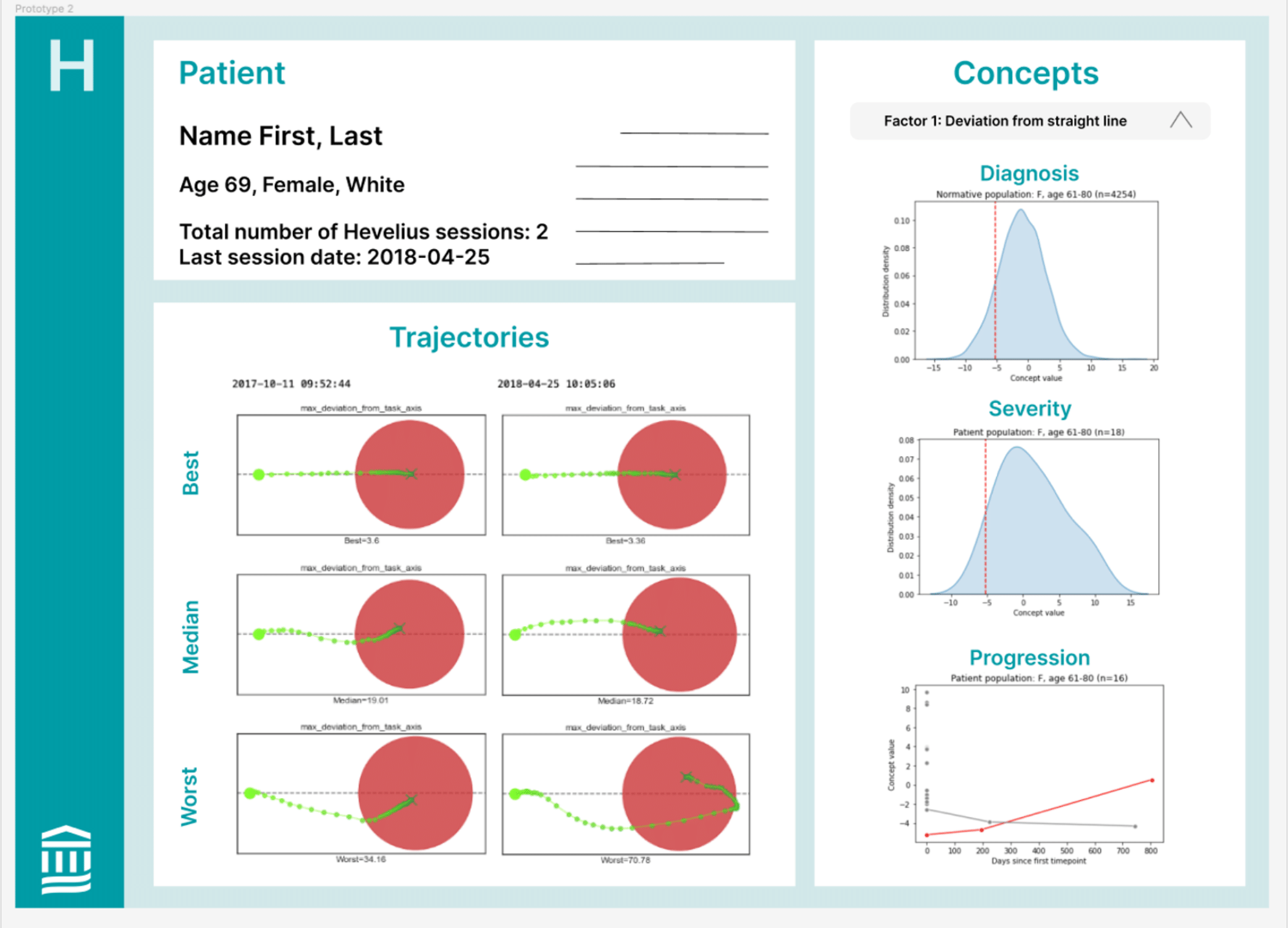}
  %\alt{A Figma frame of the first version of the user interface. It shows a one-page report with three panels. The top left panel contains the patient's basic information, such as age and gender. The bottom left panel contains the trajectory plots. The right panel contains the summary plots.}
  \caption{Design sketch for the user interface of the visualization tool. Later, during the implementation, because the amount of information and visualizations was too much to display on one screen, we opted for a single-scroll continuous page with a sidebar that allows users to jump to a section of interest.}
  \label{fig-app:dev-ui}
\end{figure}

\end{document}